# Strain-Tunable Magnetocrystalline Anisotropy in Epitaxial $Y_3Fe_5O_{12}$ Thin Films

Hailong Wang[†], Chunhui Du[†], Chris Hammel[*] and Fengyuan Yang[*]

Department of Physics, The Ohio State University, Columbus, OH, 43210, USA

[†]These authors made equal contributions to this work

[*]Emails: hammel@physics.osu.edu; fyyang@physics.osu.edu

Abstract

We demonstrate strain-tuning of magnetocrystalline anisotropy over a range of more than one thousand Gauss in epitaxial $Y_3Fe_5O_{12}$ films of excellent crystalline quality grown on lattice-mismatched $Y_3Al_5O_{12}$ substrates. Ferromagnetic resonance (FMR) measurements reveal a linear dependence of both out-of-plane and in-plane uniaxial anisotropy on the strain-induced tetragonal distortion of $Y_3Fe_5O_{12}$. Importantly, we find the spin mixing conductance $G_r$ determined from inverse spin Hall effect and FMR linewidth broadening remains large: $G_r$ = 3.33 × $10^{14}$ $\Omega^{-1}m^{-2}$ in Pt/$Y_3Fe_5O_{12}$/$Y_3Al_5O_{12}$ heterostructures, quite comparable to the value found in Pt/$Y_3Fe_5O_{12}$ grown on lattice-matched $Gd_3Ga_5O_{12}$ substrates.

Magnetocrystalline anisotropy [1-6] plays an essential role in permanent magnets, magnetic data storage, energy generation and transformation, magnetic resonance, and there is intense interest in understanding the role of magnetoelastic coupling in phonon-magnon interactions in thermal spintronics. With the growing applications of magnetic films, it is important to understand magnetocrystalline anisotropy in the presence of lattice distortion induced by epitaxial strain and the underlying magnetization-lattice coupling. Tunable magnetic anisotropy was observed in GaMnAs films at low temperatures using epitaxial strain [3], in GaMnAsP films by varying the phosphorous content [4] and in $Sr_2FeMoO_6$ epitaxial films with various strains grown on a selected set of single-crystal substrates and buffer layers [6]. Ferrimagnetic insulating $Y_3Fe_5O_{12}$ (YIG) is widely used in FMR and microwave applications as well as spin dynamics studies [7-10] due to its exceptionally low magnetic damping. Most YIG epitaxial films and single crystals are produced by liquid-phase epitaxy (LPE) with thicknesses from 100 nm to millimeters [11]. Pulsed-laser deposition (PLD) has also been used to grow epitaxial YIG thin films [12-14]. However, a systematic study of strain-dependence of magnetocrystalline anisotropy is lacking, largely due to the challenges inherent in controlling the epitaxial strain while maintaining sufficiently high crystalline quality. Strain control in high quality YIG films will allow tuning of magnetocrystalline anisotropy, which in turn determines the static and dynamic magnetization of the YIG films.

Most reported YIG epitaxial film fabrication has employed $Gd_3Ga_5O_{12}$ (GGG) substrates which has a lattice mismatch $\eta = (a_s - a_f)/a_f \times 100\%$ of 0.057% with YIG, where $a_s = 12.383$ Å and $a_f = 12.376$ Å are the lattice constants of the GGG substrate and unstrained YIG, respectively. In order to probe the magnetocrystalline anisotropy in epitaxial YIG films in response to lattice distortion, we report in this letter the growth of YIG epitaxial thin films on (001)-oriented $Y_3Al_5O_{12}$ (YAG) substrate [11, 15, 16] with $a$ = 12.003 Å ($\eta$ = -3.0%). The larger lattice mismatch results in thickness-controlled strain-induced tetragonal distortion in the YIG films, which leads to variation in their out-of-plane and in-plane magnetocrystalline anisotropy as discussed below.

We grow epitaxial YIG films with thicknesses $t$ ranging from 9.8 to 72.7 nm using a new sputtering technique [10, 17, 18] on YAG (001) substrates and determine their crystalline quality by triple-axis x-ray diffraction (XRD). Figure 1a shows $2\theta$-$\omega$ XRD scans of the YIG films of seven different thicknesses on YAG (001). The pronounced Laue oscillations observed in the 37.9-nm and 72.7-nm films indicate smooth surfaces and sharp YIG/YAG interfaces. The gradual shift of the YIG (004) peak position clearly reflects strain relaxation as the thickness of the YIG films increases from 9.8 to 72.7 nm. The lattice mismatch ($\eta$ = -3.0%, compressive) elongates the out-of-plane lattice constant $c$, resulting in a tetragonal distortion. To obtain the in-plane lattice constant $a$, we assume conservation of the unit cell volume of YIG during stain relaxation, $a = \sqrt{(12.376 \text{ Å})^3/c}$. Figure 1b and Table I show both $a$ and $c$ for the YIG films of $9.8 \leq t \leq 72.7$ nm, which exhibit a clear strain relaxation as $t$ increases, while the strain-induced tetragonal distortion $\sigma = (c - a)/a$ of the YIG films decreases from 2.05% to 0.073%.

We determine the magnetic anisotropy of our YIG films using FMR spectroscopy at radio-frequency (rf) $f$ = 9.60 GHz. A magnetic field $\boldsymbol{H}$ is applied at an angle $\theta_H$ with respect to the film normal (see inset to Fig. 2a). Figure 2a shows four representative FMR spectra for the 72.7-nm YIG film at $\theta_H$ = 0°, 30°, 50° and 90°. The resonance field $H_{res}$ is defined as the field at which the derivative of the FMR absorption crosses zero. Figure 2b shows the angular dependence of the resonance field from out-of-plane ($\theta_H$ = 0°) to in-plane ($\theta_H$ = 90°) for the 9.8, 15.0, 29.3 and 72.7-nm YIG films as the tetragonal distortion $\sigma$ varies from 2.05% to 0.073%. The magnetization can be quantitatively characterized from the total free energy density $F$ for the YIG films with tetragonal symmetry [19, 20],

$$F = -\boldsymbol{H} \cdot \boldsymbol{M} + \frac{1}{2}M\left\{4\pi M_{\text{eff}}\cos^2\theta - \frac{1}{2}H_{4\perp}\cos^4\theta - \frac{1}{8}H_{4\parallel}(3+\cos 4\phi)\sin^4\theta - H_{2\parallel}\sin^2\theta\sin^2(\phi-\frac{\pi}{4})\right\}, \quad (1)$$

where $\theta$ and $\phi$ are angles describing the orientation of the equilibrium magnetization ($\boldsymbol{M}$) (see inset to Fig. 2a). The first term in Eq. (1) is the Zeeman energy and the second term is the effective demagnetizing energy $4\pi M_{\text{eff}} = 4\pi M_s - H_{2\perp}$ which includes the shape anisotropy ($4\pi M_s$) and out-of-plane uniaxial anisotropy $H_{2\perp}$. The remaining terms are out-of-plane cubic anisotropy ($H_{4\perp}$), in-plane cubic anisotropy ($H_{4\parallel}$) and in-plane uniaxial anisotropy ($H_{2\parallel}$). We measure the magnetic hysteresis loops of the YIG films using a vibrating sample magnetometer (VSM) to obtain the saturation magnetization $M_s$. The values of $4\pi M_s$ vary from 1590 to 1850 Oe, which lie in the range of reported magnetization in YIG samples grown by LPE and PLD [11-14, 21]. The inset to Fig. 2b shows representative in-plane and out-of-plane hysteresis loops for the 37.9-nm YIG film, indicating clear magnetic shape anisotropy. Due to strain relaxation, the coercivity of our YIG films on YAG (001) ranges from 30 to 80 Oe for different thicknesses, much larger than the values of YIG films on lattice-matched GGG [13].

The equilibrium orientation ($\theta$, $\phi$) of magnetization can be obtained by minimizing the free energy, and the FMR resonance frequency $\omega$ in equilibrium is given by [19, 20, 22]

$$\left(\frac{\omega}{\gamma}\right)^2 = \frac{1}{M^2\sin^2\theta}\left[\frac{\partial^2 F}{\partial\theta^2}\frac{\partial^2 F}{\partial\phi^2} - \left(\frac{\partial^2 F}{\partial\theta\partial\phi}\right)^2\right], \quad (2)$$

where $\gamma = g\mu_B/\hbar$ is the gyromagnetic ratio. We use a numerical procedure to obtain the equilibrium angles at resonance condition [23, 24] and fit the $H_{\text{res}}$ vs. $\theta_H$ data to determine $4\pi M_{\text{eff}}$, $H_{4\perp}$, $H_{4\|}$, $H_{2\|}$, and $g$ factor. In Figure 2b, the fitting curves agree well with the experimental data which reveal a systematic variation of $4\pi M_{\text{eff}}$ for YIG films of different thicknesses. For the 9.8-nm film, $4\pi M_{\text{eff}} = 3103$ Oe while for the 72.7-nm film, $4\pi M_{\text{eff}} = 1639$ Oe, indicating that the strain induces substantial out-of-plane anisotropy. The out-of-plane uniaxial anisotropy $H_{2\perp}$ can be calculated from the values of $M_s$ and $4\pi M_{\text{eff}}$. Figure 3a shows $H_{2\perp}$ as a function of tetragonality $\sigma$ for all the YIG films on YAG; $H_{2\perp}$ varies linearly with strain. This tunability of magnetocrystalline anisotropy through lattice symmetry highlights the central result of our study: the proportionality of $H_{2\perp}$ to the tetragonal distortion of the YIG lattice over a broad range (-2.05% < $(c-a)/a$ < -0.073%),

$$H_{2\perp} = (12 \pm 64) - (55.8 \pm 5.3) \times 10^3 \times [(c-a)/a] \text{ (Oe)}.$$

Figure 3a demonstrates a fundamental relationship between magnetocrystalline anisotropy and lattice symmetry which is expected but has not been seen before in YIG films.

We also find the in-plane uniaxial anisotropy $H_{2\|}$ increases with tetragonality of the YIG lattice. Figures 2c and 2d show both the experimental data and fits to the in-plane angular dependence of $H_{\text{res}}$ for the 9.8 and 72.7 nm YIG films on YAG. Clear four-fold symmetry is observed in the 72.7-nm YIG film while superposition of two- and four-fold symmetry appears in the 9.8-nm YIG film. Based on Eqs. (1) and (2), when the in-plane anisotropy is small, the in-plane resonance condition can be expressed by [19, 20]:

$$\left(\frac{\omega}{\gamma}\right)^2 = \left\{H + H_{4\|}\cos 4\phi - H_{2\|}\cos\left(2\phi - \frac{\pi}{2}\right)\right\}$$

$$\times \left\{H + 4\pi M_{\text{eff}} + \frac{H_{4\|}(3+\cos 4\phi)}{4} + H_{2\|}\sin^2\left(\phi - \frac{\pi}{4}\right)\right\}. \qquad (3)$$

Figure 3b plots the dependence of $H_{2\parallel}$ on tetragonality $(c-a)/a$, where $H_{2\parallel}$ can be tuned from 1 to 60 Oe with the magnitude of tetragonality varying from 0.073% to 2.05%. A linear fit to Fig. 3b gives

$$H_{2\parallel} = (2 \pm 6) + (31.1 \pm 4.6) \times 10^2 \times [(c-a)/a] \text{ (Oe)}.$$

The strain-induced anisotropy arises from the magnetization-lattice coupling [25, 26] in which a change in inter-atomic distances alters the magnetic properties through spin-orbit coupling. Since $H_{2\perp}$ is more than one order of magnitude larger than $H_{2\parallel}$ in the YIG films on YAG, here we focus on the strain-induced $H_{2\perp}$. The magnetoelastic energy density is given by $F = -\sigma b$ when $M$ is along the [001] direction, where $b$ and $\sigma$ are the magnetoelastic constant and tetragonality $(c-a)/a$, respectively. Figure 3c shows the linear dependence of anisotropy energy,

$$E_{\text{ani}} = -\frac{1}{2} M H_{2\perp}, \tag{4}$$

on tetragonality for all the YIG films, from which a least squares fit gives,

$$E_{\text{ani}} = (-7.0 \pm 54.2) \times 10^2 + (40.4 \pm 4.4) \times 10^5 \times [(c-a)/a] \text{ (erg/cm}^3\text{)},$$

from which we obtain $-b = (40.4 \pm 4.4) \times 10^5$ erg/cm$^3$. The negative value of $b$ implies that the magnetic easy axis is parallel to a short axis of the tetragonal lattice. The magnetoelastic constant of YIG is somewhat smaller than but of the same order as that in double perovskite $Sr_2FeMoO_6$ films with $-b = (92.9 \pm 4.5) \times 10^5$ erg/cm$^3$ [6]. The similarity may arise because both $Y_3Fe_5O_{12}$ and $Sr_2FeMoO_6$ are $Fe^{3+}$-based ferrimagnetic oxides, while the presence of 4$d$ transition metal $Mo^{5+}$ in $Sr_2FeMoO_6$ enhances the spin-orbit coupling and, consequently, the magnetoelastic coupling. This result demonstrates the ability to tune magnetocrystalline anisotropy in thin YIG epitaxial films by substrate lattice mismatch and film thickness.

YIG is an excellent material for microwave application and spin pumping [7-10] due to its narrow linewidth and insulating nature. One fundamentally interesting question is how the strain-induced FMR linewidth broadening in YIG/YAG films will affect the spin transfer capability at YIG/Pt interface [9, 10]. It is believed that the FMR linewidth largely determines the quality of YIG films and interfacial spin mixing conductance $G_r$ in YIG/normal-metal bilayers. Here, we report cavity FMR spin pumping measurements in Pt/YIG/YAG . The FMR peak to peak linewidth $\Delta H$ is 83.9Oe for the 72.7 nm YIG film on YAG. Figure 4a shows the $V_{ISHE}$ vs. $H$ spectra for Pt( 5 nm)/YIG(72.7 nm)/YAG with an in-plane DC field ***H*** at $P_{rf}$ = 200 mW. The ISHE signal is 123 μV which, although smaller than our previously reported mV $V_{ISHE}$ for Pt/YIG on GGG [10], is still large for Pt/YIG system. Figure 4b shows the FMR derivative absorption spectra of a single 72.7-nm YIG film and a Pt(5 nm)/YIG(72.7 nm) bilayer on YAG. The real part of interfacial spin mixing conductance $G_r$ can be determined from [27, 28],

$$G_r = \frac{e^2}{h} \frac{2\sqrt{3}\pi M_s \gamma t_F}{g\mu_B \omega} (\Delta H_{Pt/YIG} - \Delta H_{YIG}) \qquad (5)$$

where $\gamma$, $g$, $\mu_B$ and $t_F$ are the gyromagnetic ratio, $g$ factor, Bohr magnetron and thickness of YIG film, respectively. Using Eq. (5) and the linewidths from Fig. 4b, we obtain the spin mixing conductance $(3.33 \pm 0.15) \times 10^{14}$ $\Omega^{-1}m^{-2}$ for Pt/YIG on YAG, which is slightly smaller but comparable to the values of $3.73 \times 10^{14}$ and $4.56 \times 10^{14}$ $\Omega^{-1}m^{-2}$ for Pt/YIG bilayers on GGG [10]. This indicates that the larger FMR linewidth for YIG films grown on YAG essentially does not change the effective spin angular momentum transfer capability across the Pt/YIG interface. One possible explanation is that the strain-induced inhomogeneity mostly exists in the bulk of the YIG film and the Pt/YIG interface remains high quality.

The tunable magnetocrystalline anisotropy in strained YIG thin films with a clear linear dependence on the tetragonal distortion of YIG lattice allows for fundamental understanding of magnetization-lattice coupling in this important magnetic material and enables potential microwave and spin-electronic applications via control of the lattice symmetry. This behavior points towards potential strain engineering of YIG epitaxial films, for example, with lateral modulation of strain to tune the magnetic resonance characteristics and to design microwave heterostructures for novel applications.

**Acknowledgements**

This work is supported by the Center for Emergent Materials at the Ohio State University, a NSF Materials Research Science and Engineering Center (DMR-0820414) (HLW and FYY) and by the Department of Energy through grant DE-FG02-03ER46054 (PCH). Partial support is provided by Lake Shore Cryogenics Inc. (CHD) and the NanoSystems Laboratory at the Ohio State University

**Figure Captions:**

**Figure 1.** (a) Semi-log $2\theta$-$\omega$ XRD scans of YIG films of thickness $t$ = 9.8, 12.4, 15.0, 19.5, 29.3, 37.9 and 72.7 nm grown epitaxially on YAG (001) substrates. The arrows indicate the positions of the YIG (004) peak. The satellite peaks in the scans of 37.9 and 72.7 nm YIG films are the Laue oscillations. (b) Thickness dependence of the in-plane (blue open squares) lattice constant $a$ and out-of-plane (red solid circles) lattice constant $c$ of the YIG films on YAG. The horizontal dashed line represents the bulk lattice constant $a$ = 12.376 Å of YIG.

**Figure 2.** (a) Room-temperature FMR derivative spectra for a 72.7 nm YIG film on YAG (001) at $\theta_H$ = 0°, 30°, 50°, and 90°. Inset: coordinate system used for FMR measurements and analysis. (b) Out-of-plane angular dependence ($\theta_H$) of the resonance fields ($H_{res}$) for the 9.8, 15.0, 29.3, and 72.7 nm YIG films. The fitting (solid curves) was performed using Eqs. (1) and (2) to obtain $4\pi M_{eff}$, from which $H_{2\perp}$ was determined for each film. Inset: in-plane (blue) and out-of-plane (red) magnetic hysteresis loops of a 37.9-nm thick YIG film. In-plane angular dependence ($\phi_H$) of $H_{res}$ for the (c) 9.8 nm and (d) 72.7 nm YIG films.

**Figure 3.** (a) Out-of-plane uniaxial anisotropy field $H_{2\perp}$, (b) in-plane anisotropy field $H_{2\parallel}$ and (c) (c) out-of-plane anisotropy energy $E_{ani}$ as a function of the tetragonal distortion $(c-a)/a$ of the YIG films.

**Figure 4.** (a) $V_{ISHE}$ vs. $H$ spectra at $\theta_H$ = 90° and 270° using $P_{rf}$ = 200 mW for a Pt(5 nm)/YIG(72.7 nm) bilayer. Inset: FMR spin pumping experimental geometry. (b) FMR derivative absorption spectra of the 72.7-nm thick YIG film before (red) and after (blue) the deposition of a 5-nm Pt layer.

**Table I**. Structural and magnetic parameters of YIG epitaxial films with thickness $9.8 \leq t \leq 72.7$ nm on YAG (001).

| $t$ (nm) | $a$ (Å) | $c$ (Å) | $(c-a)/a$ | $H_{2\perp}$ (Oe) | $E_{ani}$ (erg/cm$^3$) | $H_{2\parallel}$ (Oe) | $H_{4\parallel}$ (Oe) |
|---|---|---|---|---|---|---|---|
| 9.8 | 12.293 | 12.545 | 2.05% | $-1.25 \times 10^3$ | $9.22 \times 10^4$ | 60.4 | 42.0 |
| 12.4 | 12.308 | 12.513 | 1.66% | -902 | $6.30 \times 10^4$ | 48.7 | 58.6 |
| 15.0 | 12.318 | 12.493 | 1.43% | -701 | $4.57 \times 10^4$ | 52.1 | 66.8 |
| 19.5 | 12.334 | 12.460 | 1.03% | -543 | $3.77 \times 10^4$ | 17.9 | 17.9 |
| 29.3 | 12.354 | 12.420 | 0.53% | -445 | $2.91 \times 10^4$ | 23.9 | 18.0 |
| 37.9 | 12.363 | 12.402 | 0.31% | -139 | $1.00 \times 10^4$ | 2.75 | 25.9 |
| 72.7 | 12.373 | 12.382 | 0.073% | -49 | $3.10 \times 10^3$ | 0.941 | 25.6 |

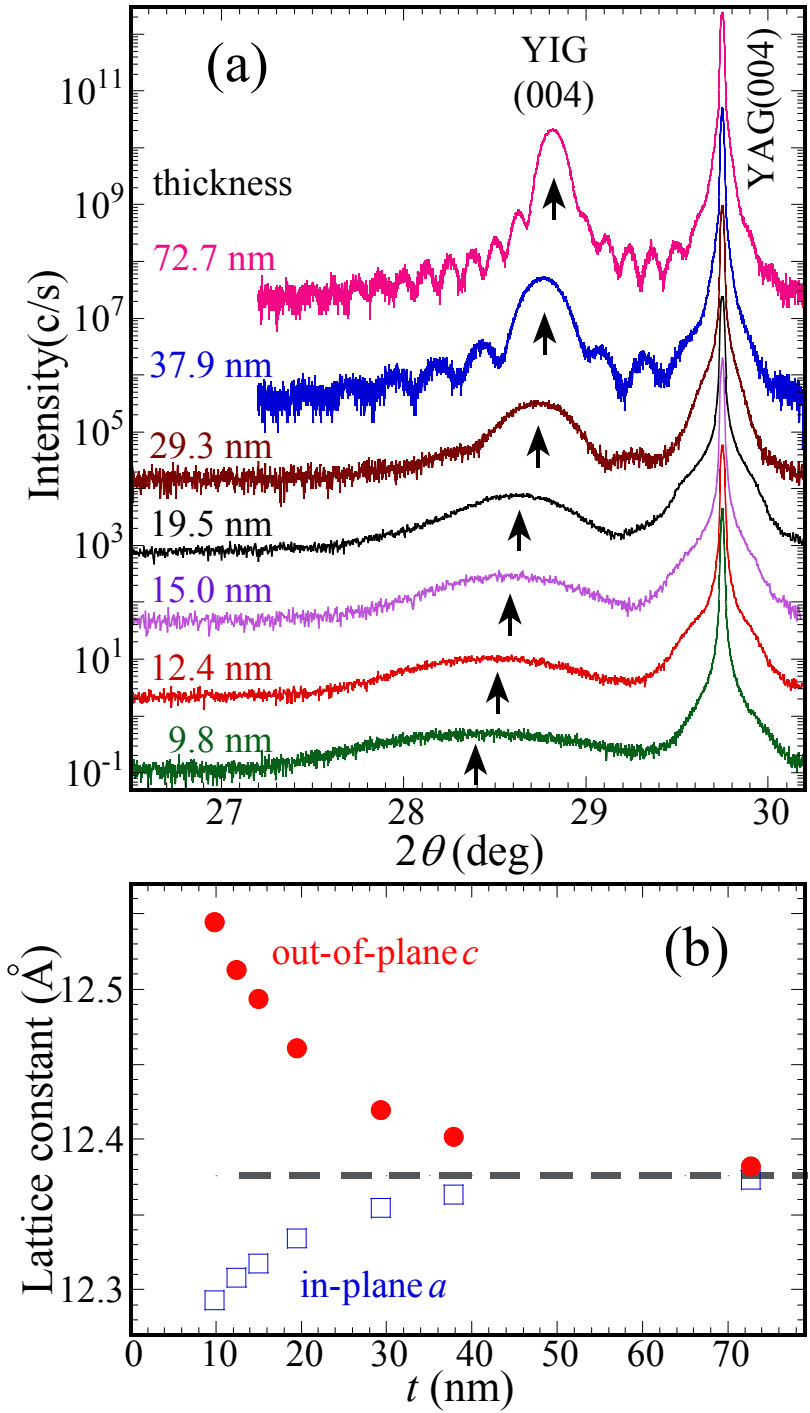

**Figure 1.**

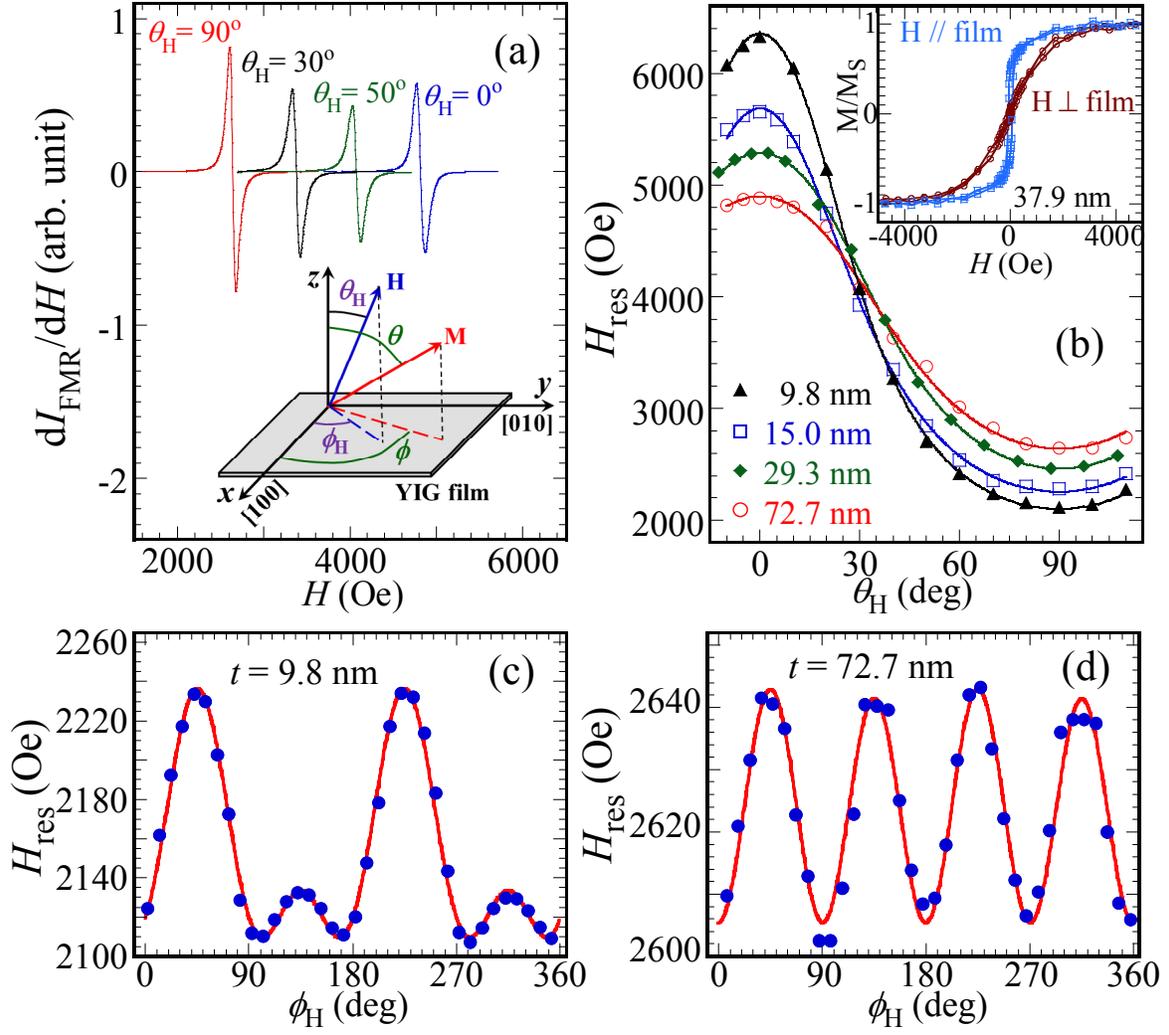

**Figure 2.**

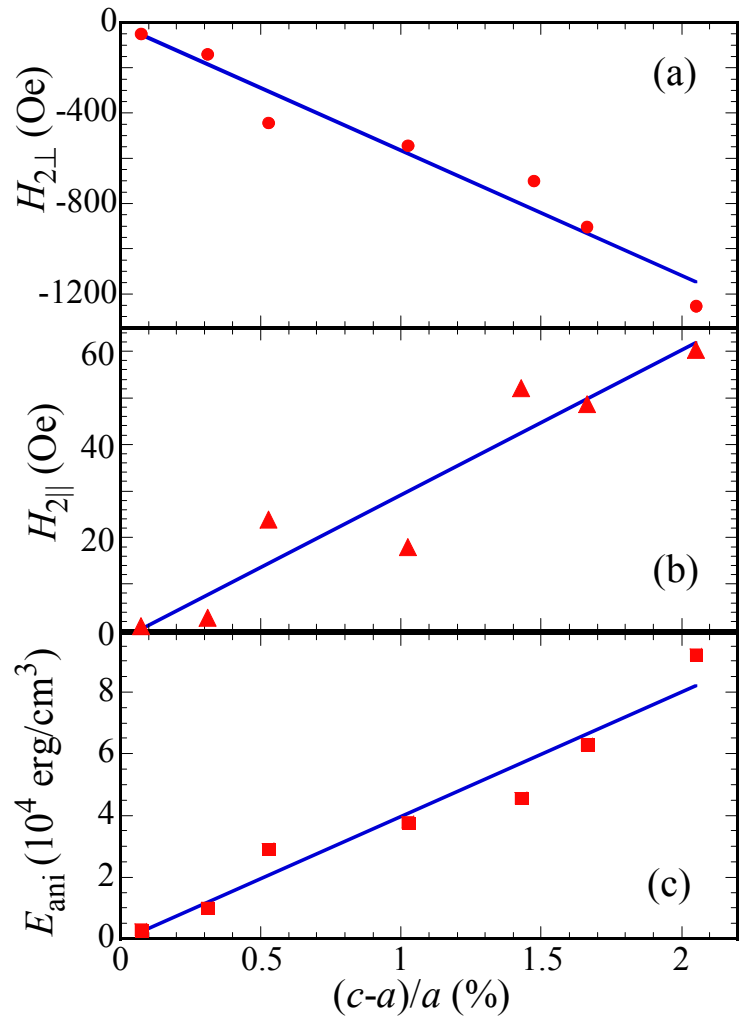

**Figure 3**.

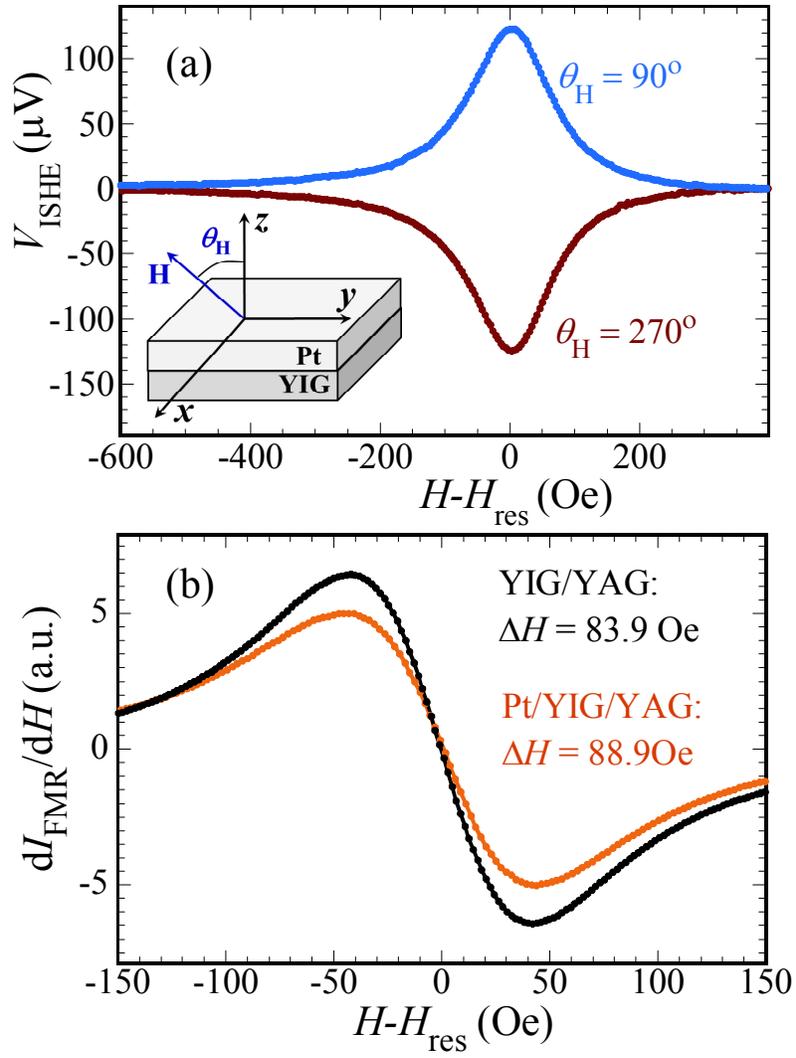

**Figure 4.**